\documentclass[12pt,a4paper]{article}
\usepackage{graphicx}
\usepackage[T1]{fontenc}
\usepackage[utf8]{inputenc}
\usepackage{textcomp}
\usepackage[sc,osf]{mathpazo}
\usepackage{a4wide}  
\usepackage{latexsym,amsthm,amsfonts,amsmath,mathrsfs,amssymb}
\usepackage{dsfont}
\usepackage{accents}
\usepackage[nosort]{cite}
\usepackage{booktabs} 
\usepackage[unicode,implicit]{hyperref}
\hypersetup{%
  pdftitle    = {A note on the calculation of the Komar integral in the
    Lorentzian Taub-NUT spacetime}
  pdfkeywords = {gravity,  conserved quantities, Taub-NUT solution, Misner strings},
  pdfauthor   = {Gabriele Barbagallo, Jos\'e Luis V. Cerdeira, Carmen G\'omez-Fayr\'en and Tom\'as Ort\'{\i}n},
  plainpages  = true,
  colorlinks  = true,
  citecolor   = blue,
  urlcolor    = red,
  linkcolor   = black
}
\newcommand{\hepth}[1]{{\tt
\href{http://www.arXiv.org/abs/hep-th/#1}{hep-th/#1}}}
\newcommand{\grqc}[1]{{\tt
\href{http://www.arXiv.org/abs/gr-qc/#1}{gr-qc/#1}}}

\newcommand{\arxiv}[1]{{\tt arXiv:\href{http://www.arXiv.org/abs/#1}{#1}}}
\newcommand{\doi}[1]{{\tt DOI:\href{http://dx.doi.org/#1}{#1}}}

\allowdisplaybreaks

\begin{document}

\begin{flushright}
\small
IFT-UAM/CSIC-25-050\\
May 20\textsuperscript{th}, 2025\\
\normalsize
\end{flushright}

\vspace{.2cm}

\begin{center}

  {\Large {\bf A note on the calculation of the Komar integral\\[.5cm] in the
    Lorentzian Taub-NUT spacetime}}
 
\vspace{1cm}

\renewcommand{\thefootnote}{\alph{footnote}}

{\sl Gabriele Barbagallo,$^{1,}$}\footnote{Email: {\tt gabriele.barbagallo[at]estudiante.uam.es}}
{\sl Jos\'e Luis V.~Cerdeira}$^{2,}$\footnote{Email: {\tt jose.verez-fraguela[at]estudiante.uam.es}}
{\sl Carmen G\'omez-Fayr\'en,$^{1,}$}\footnote{Email: {\tt carmen.gomezfayren[at]csic.es}}
{\sl and Tom\'{a}s Ort\'{\i}n}$^{1,}$\footnote{Email: {\tt Tomas.Ortin[at]csic.es}}

\setcounter{footnote}{0}
\renewcommand{\thefootnote}{\arabic{footnote}}
\vspace{1cm}

${}^{1}${\it\small Instituto de F\'{\i}sica Te\'orica UAM/CSIC\\
C/ Nicol\'as Cabrera, 13--15,  C.U.~Cantoblanco, E-28049 Madrid, Spain}

\vspace{0.2cm}

${}^{2}${\it\small Instituto de F\'{\i}sica Corpuscular (IFIC), University of
  Valencia-CSIC,\\
Parc Cient\'{\i}fic UV, C/ Catedr\'atico Jos\'e Beltr\'an 2, E-46980 Paterna, Spain}

\vspace{3cm}

{\bf Abstract}
\end{center}
\begin{quotation}
  {\small It has recently been shown that one can derive consistent
    thermodynamical expressions in the Lorentzian Taub--NUT spacetime keeping
    the Misner-string singularities and taking into account their
    contributions in the Komar integrals. We show how the same results are
    obtained when the Mister-string singularities are removed by using
    Misner's procedure because, even though the complete spacetime has no such
    singularities anymore, they are unavoidable in all spacelike hypersurfaces
    which are used in the Komar integrals.  Different choices of hypersurfaces
    may contain different strings and lead to different physics, though.
  }
\end{quotation}

\newpage
\pagestyle{plain}

\noindent
\textbf{1. Introduction.} The Lorentzian Taub-NUT solution of the vacuum
Einstein equations \cite{Taub:1950ez,Newman:1963yy} can be seen as a
relatively simple generalization of the Schwarzschild solution of mass $M$
that includes a new parameter, the \textit{NUT charge} $N$, that can be
interpreted as a ``magnetic mass'' (if we view the standard ADM mass as the
``electric mass''). This solution can be written in the form
\begin{equation}
\label{eq:TNsolution1}
ds^{2}
=
\lambda(r) (dt +A )^{2} 
-\lambda^{-1}(r)dr^{2} -\left(r^{2}+N^{2} \right)d\Omega_{(2)}^{2}\,,
\end{equation}
where
\begin{subequations}
\label{eq:TNsolution2}
\begin{align}
\lambda(r)
  & =
\frac{(r-r_{+})(r-r_{-})}{r^{2}+N^{2}}\,,
  \\
  & \nonumber \\
  \label{eq:1formA}
  A
  & =
    2N\cos{\theta}\,d\varphi\,,
  \\
  & \nonumber  \\
  d\Omega_{(2)}^{2}
  & =
    d\theta^{2}+\sin^{2}{\theta}d\varphi^{2}\,,
\end{align}
\end{subequations}
and where we have defined 
\begin{equation}
r_{\pm}
=
M\pm r_{0}\,,
\hspace{1cm} 
r_{0}^{2} = M^{2}+N^{2}.  
\end{equation}

The presence of this intriguing parameter is one of the most interesting
aspects of this solution.  The parallelism between the NUT charge and the
magnetic charge of the Dirac monopole is reinforced by the presence of
\textit{Misner-string singularities} in the Taub-NUT solution, which bear a
strong similarity to the Dirac-string singularities of the electromagnetic
field of the Dirac magnetic monopole \cite{Dirac:1931kp}, as we are going to
see.

First of all, notice that the the 1-form $A$ of the Taub--NUT solution
Eq.~(\ref{eq:1formA}) is identical to the field of the Dirac monopole in some
gauge. Coordinate transformations of the form
\begin{equation}
  t\to t+\chi(x)\,,
  \hspace{1cm}
  \partial_{t}\chi =0\,,
\end{equation}
are equivalent to  transformations of $A$ of the form
\begin{equation}
A\to A+d\chi\,,  
\end{equation}
and to gauge transformations of the Dirac monopole field. In order to see
where the string singularities lie, it is convenient to perform a gauge
transformation $t\to t+2Ns\varphi$, where the parameter $s$ can take the
values $-1,0,+1$. The 1-form $A$ takes the form
\begin{equation}
  \label{eq:1formAs}
  A
  =
  2N(\cos{\theta}+s)\, d\varphi\,,
\end{equation}
or
\begin{equation}
  \label{eq:1formAsCartesian}
  A
  =
  2N(\cos{\theta}+s)\, d\varphi
  =
  2N\frac{z+sr}{r(r^{2}-z^{2})}(xdy-ydx)\,,
\end{equation}
in Cartesian coordinates.  The denominator goes to zero quadratically at
$r=|z|$, (the whole $z$ axis, $x=y=0$). Thus, the 1-form will have a
singularity when $z+s|z|\neq 0$, that is
\begin{equation}
  \begin{aligned}
    s & = 0\,,\hspace{1cm} & \text{the whole $z$ axis} \\
    s & = +1\,,\hspace{1cm} & \text{the $z>0$ semiaxis} \\
    s & = -1\,,\hspace{1cm} & \text{the $z<0$ semiaxis} \\
  \end{aligned}
\end{equation}
These singularities of the electromagnetic field are the Dirac strings. For
different values of $s$ we obtain different field configurations of the
electromagnetic field. They are related by large gauge transformations that
modify the asymptotic behaviour and, therefore, cannot be seen as physically
equivalent. However, since they should be undetectable by electrically charged
particles whose charges obey the Dirac quantization condition
\cite{Dirac:1931kp}, they have long been considered unphysical. The fact that,
as we are going to see, one can obtain stringless solutions with similar
properties \cite{Wu:1975es} seems to support this point of view.

In the case of the Taub--NUT metric, we find something similar: after
performing the above transformation (now, a general coordinate
transformation), we find that
\begin{equation}
  g^{tt}
  =
\frac{1}{\lambda}-\frac{4N^{2}(\cos\theta+s)^{2}}{(r^{2}+N^{2})\sin^{2}\theta}\,,
\end{equation}
and
\begin{equation}
  \frac{(\cos{\theta} +s)^{2}}{\sin^{2}{\theta}}
  =
  \left\{
    \begin{array}{lll}
    \cot^{2}{\theta}\,,& \text{for $s=0$,} \hspace{.5cm} & \text{singular at
                                                             $\theta =0,\pi
                                                             \Rightarrow
                                             \forall z$}
      \\
   \cot^{2}{(\theta/2)}\,,& \text{for $s=+1$,} & \text{singular at
                                                             $\theta = 0
                                                                \Rightarrow z>0$}
                                                                \\
   \tan^{2}{(\theta/2)}\,,& \text{for $s=-1$,} & \text{singular at
                                                             $\theta =\pi
                                                             \Rightarrow z<0$}
    \end{array}
  \right.
\end{equation}

Misner showed in Ref.~\cite{Misner:1963fr} that one can construct a solution
in which these singularities are absent by using in the $z>0$ region a
Taub--NUT solution whose  Misner string lies in the negative $z$-axis
($s=-1$)\footnote{Strictly speaking, all the coordinates in this piece of the
  should bear a $(+)$ label. However, the transformation between the
  coordinates in the two patches that we are going to use will be trivial for
  $r,\theta$ and $\varphi$, and we do not write this superscript for the sake
  of simplicity. A similar comment applies to the $(-)$ piece of the
  solution.}
\begin{equation}
\label{eq:TN+}
ds^{2}
 =
\lambda(r) \left[dt^{(+)} +2N(\cos{\theta}-1)\,d\varphi \right]^{2} 
-\lambda^{-1}(r)dr^{2} -\left(r^{2}+N^{2} \right)d\Omega_{(2)}^{2}\,,
\end{equation}
in the $z<0$ region, a Taub--NUT solution whose  Misner string lies in
the positive $z$-axis ($s=+1$),
\begin{equation}
\label{eq:TN-}
ds^{2}
 =
\lambda(r) \left[dt^{(-)} +2N(\cos{\theta}+1)\,d\varphi \right]^{2} 
-\lambda^{-1}(r)dr^{2} -\left(r^{2}+N^{2} \right)d\Omega_{(2)}^{2}\,,
\end{equation}
and gluing them across the equatorial plane $z=0$, \textit{i.e.}~relating them
by a coordinate transformation in the overlap:
\begin{equation}
  \label{eq:t+versust-}
  t^{(+)}
  =
  t^{(-)}+4N\varphi\,,
\end{equation}

Since the period of the angular coordinate $\varphi$ is $2\pi$ and we must
identify $\varphi\sim \varphi+2\pi$, the time coordinate in both patches must
also be periodic with period $8\pi N$
\begin{equation}
  t^{(\pm)}
  \sim
  t^{(\pm)} +8\pi N.
\end{equation}

In the resulting solution, the time coordinate parametrizes a S$^{1}$ which is
non-trivially fibered over the S$^{2}$ parametrized by the coordinates
$\theta$ and $\varphi$ at each value of the radial coordinate $r$.  The new
solution is locally identical to the stringless regions of the original
Taub--NUT solution Eq.~(\ref{eq:TNsolution1}). The information contained in
the missing Misner strings is now contained in the non-trivial topology.

In Ref.~\cite{Wu:1975es} Wu and Yang showed how to construct a non-trivial
U$(1)$ ($\sim$ S$^{1}$) bundle over S$^{2}$ locally identical to the
stringless regions of the Dirac magnetic monopole. This construction is
essentially identical to Misner's.

Evidently, the periodic identification of the time coordinate comes with its
own problems, like the presence of closed timelike curves, to start
with. Another problem of special interest to us is that the Wick-rotated
solution has a conical singularity at the location of the event horizon
$r=r_{+}$ unless the Euclidean time has a period of $\beta = 2\pi/\kappa$,
$\kappa$ being the horizon's surface gravity. The consistency condition
$\beta=8\pi N$ constrains the parameter space of the solution, leaving us with
a single free parameter and leading to the loss of full cohomogeneity in the
first law of black-hole thermodynamics obtained through the Euclidean approach.

In Refs.~\cite{Clement:2015cxa,Clement:2016mll} it was shown that, perhaps,
there is no need to remove the Misner-string singularities because they are
rather mild and the spacetime is geodesically complete (the singularities are
transparent for free-falling observers) and essentially free of causal
pathologies. If the strings are kept, the problems that arise in the Euclidean
approach to black-hole thermodynamics disappear \cite{Hennigar:2019ive} and
one can recover the same results in the Lorentzian setting if one takes into
account the contributions of the strings \cite{Bordo:2019tyh}.\footnote{The
  extension to electrovac spacetimes was made in Ref.~\cite{Clement:2019ghi}.}

\vspace{.3cm}
\noindent
\textbf{2. The Lorentzian approach to Taub--NUT thermodynamics.} In the
Lorentzian approach to black-hole thermodynamics both the Smarr formula and
the first law can be derived by integrating the exterior derivatives of
certain on-shell closed 2-forms over spacelike hypersurfaces with two
boundaries: one at a section of the horizon (usually the bifurcation sphere
$\mathcal{BH}$) and the 2-sphere at spatial infinity S$^{2}_{\infty}$. Stokes
theorem leads to an identity relating the integrals of the 2-forms at both
boundaries which becomes the Smarr formula
\cite{Bardeen:1973gs,Carter:1973rla} or the first law \cite{Wald:1993nt}.

The 2-form whose (vanishing) exterior derivative one has to integrate in order
to obtain the Smarr formula is the Komar charge \cite{Komar:1958wp} (General
Relativity without matter) or generalizations thereof
\cite{Magnon:1985sc,Bazanski:1990qd,Kastor:2010gq}\footnote{See also
  Ref.~\cite{Ballesteros:2024prz} and references therein for more recent works
  on generalized Komar charges and Smarr formulas.} associated to the Killing
vector $k$ that becomes null over the event horizon, $\mathbf{K}[k]$. If
$\Sigma^{3}$ is a spacelike hypersurface such that
$\partial \Sigma^{3} = \mathcal{BH} \cup $S$^{2}_{\infty}$, taking into
account the different orientation of these two boundaries
\begin{equation}
  0 = \int_{\Sigma^{3}}d\mathbf{K}[k]
  =
  \int_{\mathcal{BH}} \mathbf{K}[k] - \int_{\text{S}^{2}_{\infty}} \mathbf{K}[k]\,. 
\end{equation}

The integral at infinity can be expressed in terms of conserved charges, which
are defined asymptotically, while the integral over the bifurcation sphere can
be expressed in terms of properties of the horizon: temperature, entropy,
potentials etc. The identity of these two results is the Smarr formula.

In the case of the Taub--NUT solution with Misner strings,  the boundary of
the spacelike surface $\Sigma^{3}$ contains one additional boundary per string
and
\begin{equation}
  0 = \int_{\Sigma^{3}}d\mathbf{K}[k]
  =
  \int_{\mathcal{BH}} \mathbf{K}[k] - \int_{\text{S}^{2}_{\infty}}
  \mathbf{K}[k]
  +\sum_{\rm strings}\int_{\rm string} \mathbf{K}[k]\,.
\end{equation}

\begin{figure}[ht!]
\begin{center}
\includegraphics[scale=0.9,trim=0 350 0 130,clip]{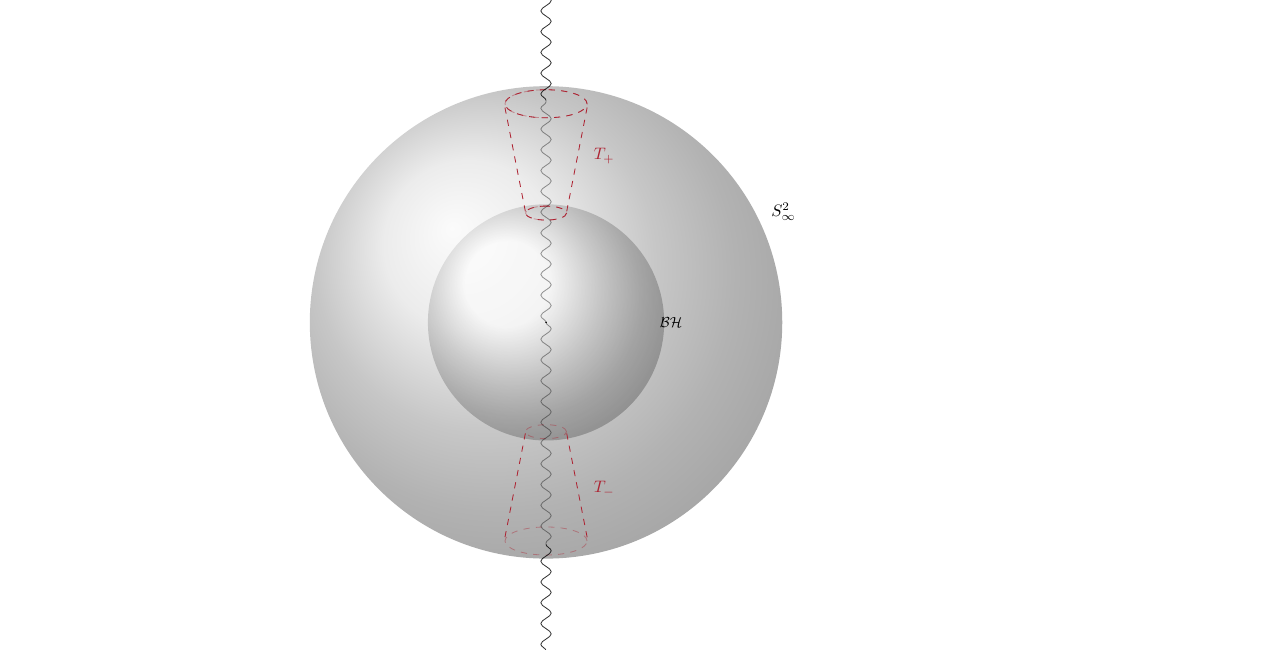}
\caption{This figure represents schematically the region over which we
  integrate $d\mathbf{K}[k]$, \textit{i.e.}~the region contained inside the
  boundaries S$^{2}_{\infty}$ (the 2-sphere at spatial infinity),
  $\mathcal{BH}$ (the bifurcation sphere) and the two cones T$_{+}$ and
  T$_{-}$ surrounding the Misner-string singularities (when they are
  present).}
\label{fig:Drawing}
\end{center}
\end{figure}

In practice (see Fig~\ref{fig:Drawing}) the integrals over the strings are
computed as integrals over cones (surfaces of fixed $t$ and $\theta$ are
actually cones rather than tubes) from $r=r_{+}$ (the horizon) to some finite
value of $r$, the integral over the bifurcation sphere now must exclude the
regions inside the cones that surround the intersection between the strings
and the sphere\footnote{This is $\theta=\epsilon$ if there is a string along
  the $z>0$ semiaxis and/or $\theta=\pi-\epsilon$ if there is a string along
  the $z<0$ semiaxis. $\epsilon$ is assumed to be positive and small.} and the
integral over the sphere at infinity is computed as an integral over a sphere
of finite radius $r$ excluding the regions that surround the intersection
between the strings and the sphere. Then, the final result is obtained by
taking the limits $r\to \infty$ and $\epsilon\to 0$.

The first two integrals in the above expression give usual results:
\begin{subequations}
  \begin{align}
    \int_{\mathcal{BH}} \mathbf{K}[k]
    & =
      ST\,,
    \\
    & \nonumber \\
    \int_{\text{S}^{2}_{\infty}} \mathbf{K}[k]
    & =
      M/2\,,
  \end{align}
\end{subequations}
where $S$ is the Bekenstein--Hawking entropy ($A/4G_{N}^{(4)}$, where $A$ is
the area of any section of the horizon), $T$ is the temperature
($\kappa/(2\pi)$ where $\kappa$ is the surface gravity of the horizon) and $M$
is the ADM mass. There is dependence on the NUT charge hidden in $S$ and $T$
and the standard Smarr formula $M=2ST$ is not satisfied.

The integrals over the strings in the $z>0$ and $z<0$ regions provide the
missing terms and can be written in the form
\begin{equation}
  \int_{\text{string}_{\pm}} \mathbf{K}[k]
  =
  \psi_{\pm} \mathcal{N}_{\pm}\,,
\end{equation}
where $\psi_{\pm}$ play the role of chemical potentials and coincide with the
surface gravities of the strings and $\mathcal{N}_{\pm}$, the thermodynamical
variables conjugate to them, are related but not equal to the NUT charge $N$.

With these additional contributions, the Smarr formula takes the general form
\cite{Bordo:2019tyh}
\begin{equation}
  \label{eq:SmarrNUT}
  M
  =
  2\left(TS +\psi_{+}\mathcal{N}_{+} +\psi_{-}\mathcal{N}_{-}\right)\,,
\end{equation}
and it is identically satisfied by the Taub--NUT solutions with $s=\pm 1,0$,
for which $\mathcal{N}_{\mp}=0$ and $\mathcal{N}_{\pm}\neq 0$, respectively.

\vspace{.3cm}
\noindent
\textbf{3. The Lorentzian approach to the stringless Taub--NUT spacetime.}
What happens when one uses Misner's construction to remove the string
singularities? The Komar integrals associated to the strings no longer
contribute to the Smarr formula but one may expect the integrals over the
spheres to give additional contributions owing to the fact that they have to
be computed for the northern ($z>0$) and southern hemispheres ($z<0$)
separately, but it can be seen by direct calculation that this expectation is
not fulfilled and one apparently gets the inconsistent result $M=2ST$.

At this point we should remember that we must integrate\footnote{We indicate
  with $\doteq$ identities which may only be satified on-shell, with $\,\d{=}$
  identities which are noly satisfied when the gauge parameter (here, a vector
  field) is \textit{reducibility} of \textit{Killing parameter}
  \cite{Barnich:2001jy,Barnich:2003xg} (here, a Killing vector) and with
  $\doteqdot$ identities which may only be satisfied when both conditions are
  met.}  $d\mathbf{K}[k]\doteqdot 0$ over a spacelike hypersurface ($t=0$,
say) and we must pay attention to the definition of this hypersurface in the
stringless spacetime, which has two coordinate patches and two time
coordinates. The choice $t^{(+)}=0$ implies, according to
Eq.~(\ref{eq:t+versust-}), $t^{(-)} = -4 N \varphi$ and the induced metric of
the hypersurface in both hemispheres is the same metric one gets by setting
$t=0$ in the solution
\begin{equation}
ds^{2}
 =
\lambda(r) \left[dt +2N(\cos{\theta}-1)\,d\varphi \right]^{2} 
-\lambda^{-1}(r)dr^{2} -\left(r^{2}+N^{2} \right)d\Omega_{(2)}^{2}\,,
\end{equation}
which has a Misner-string singularity in the $z<0$ axis.

In the same way, we find a Misner-string singularity along the $z>0$ semiaxis
in the hypersurface defined by $t^{(-)}=0$ ($t^{(+)}=4N\varphi$) and, in the
hypersurface $t^{(+)}=2N\varphi$ ($t^{(-)}=-2N \varphi$), there is a
Misner-string singularity along both semiaxes. These three choices of
spacelike hypersurfaces are obviously related to the three values of $s$ we
have considered here, although there can be more.\footnote{Notice that the
  choice of spacelike hypersurface is a choice of gauge for the connection
  field $A$. There is no globally regular gauge choice.}

Thus, although the stringless Taub--NUT solution is completely regular outside
the horizon, its spacelike hypersurfaces exhibit Misner-string singularities
that must be taken into account exactly in the same fashion they are taken
into account when the singularities are present in the full solution. Thus,
integrating $d\mathbf{K}[k]\doteqdot 0$ over these hypersurfaces one has to do
exactly the same calculations as in Ref.~\cite{Bordo:2019tyh} and one gets
results that fit in the general Smarr formula Eq.~(\ref{eq:SmarrNUT}). There
is, however an important difference: in the solutions with Misner-string
singularities one gets different values of the charges and potentials for
different values of $s$. This is acceptable because they are physically
inequivalent solutions with different string singularities. However, obtaining
different values for different choices of hypersurface in the globally
regular, stringless solution indicates that these hypersurfaces are physically
inequivalent and observers associated to them will experience different
physical phenomena.

The Euclidean approach does not make use of hypersurfaces and the
Misner-string singularities associated to them seem to be totally absent. The
problems found in the thermodynamical description of the stringless Taub--NUT
solution are unaffected by our observations.

\vspace{.3cm}
\noindent
\textbf{Acknowledgments} The authors would like to thank R.~Hennigar for
reading the manuscript and useful comments.  The work of GB, CG-F, TO and JLVC
has been supported in part by the MCI, AEI, FEDER (UE) grants
PID2021-125700NB-C21 (``Gravity, Supergravity and Superstrings'' (GRASS)) and
IFT Centro de Excelencia Severo Ochoa CEX2020-001007-S.  The work of GB has
been supported by the fellowship CEX2020-001007-S-20-5. The work of CG-F was
supported by the MU grant FPU21/02222. The work of JLVC has been supported by
the CSIC JAE-INTRO grant JAEINT-24-02806. TO wishes to thank M.M.~Fern\'andez
for her permanent support.

\appendix


\end{document}